\documentclass[fleqn,usenatbib]{mnras}

\usepackage{newtxtext,newtxmath}

\usepackage[T1]{fontenc}

\DeclareRobustCommand{\VAN}[3]{#2}
\let\VANthebibliography\thebibliography
\def\thebibliography{\DeclareRobustCommand{\VAN}[3]{##3}\VANthebibliography}

\usepackage{graphicx}	
\usepackage{amsmath}	
\usepackage{bm}
\usepackage[ISO]{diffcoeff}

\title[Magnetic field perturbations in Uranus]{Linking Uranus' temperature profile to wind--induced magnetic fields}

\author[]{
Deniz Soyuer,\thanks{E-mail: deniz.soyuer@uzh.ch}
and Ravit Helled
\\
Center for Theoretical Astrophysics and Cosmology, Institute for Computational Science, University of Zurich, Winterthurerstrasse 190, CH-8057 Zurich,\\
Switzerland\\
}

\date{Accepted 2021 August 3. Received 2021 August 3; in original form 2021 June 10.}

\pubyear{2021}

\begin{document}
\label{firstpage}
\pagerange{\pageref{firstpage}--\pageref{lastpage}}
\maketitle

\begin{abstract}
The low luminosity of Uranus is still a puzzling phenomenon and has key implications for the thermal and compositional gradients within the planet.
Recent studies have shown that planetary volatiles become ionically conducting under conditions that are present in the ice giants.
Rapidly growing  electrical conductivity with increasing depth would couple zonal flows to the background magnetic field in the planets, inducing poloidal and toroidal field perturbations   $\bm{B}^{\omega} = \bm{B}^{\omega}_P + \bm{B}^{\omega}_T$ via the $\omega$--effect. Toroidal perturbations $\bm{B}^{\omega}_T$ are expected to diffuse downwards and produce poloidal fields $\bm{B}^{\alpha}_P$ through turbulent convection via the $\alpha$--effect, comparable in strength to those of the $\omega$--effect; $\bm{B}^{\omega}_P$. 
To estimate the strength of poloidal field perturbations for various Uranus models in the literature, we generate wind decay profiles based on Ohmic dissipation constraints assuming an ionically conducting H$_2$--He--H$_2$O interior.
Due to the higher  metallicities in outer regions of hot Uranus models, zonal winds need to decay to $\sim$0.1\% of their surface values in the outer 1\% of Uranus to admit decay solutions in the Ohmic framework. 
Our estimates suggest that colder Uranus models could potentially have poloidal field perturbations that reach up to $\mathcal{O}(0.1)$ of the background magnetic field in the most extreme case. The possible existence of poloidal field perturbations spatially correlated with Uranus' zonal flows could be used to constrain Uranus' interior structure, and presents a further case for the \textit{in situ} exploration of Uranus. 
\end{abstract}




\section{Introduction}
The solar system's giant planets provide us with an exceptional opportunity for studying the physics of high–pressure, rotating systems with density and compositional variations. 
Among them, Jupiter and Saturn have been relatively well-studied compared to Uranus and Neptune, which remain the least explored solar system planets to this day. 
The scientific potential of space missions to Uranus and Neptune is currently being thoroughly assessed by the planetary science  community \citep[see e.g.][]{exp_hof,exp_helled, exp_dahl, fletcher,  exp_fle,  whitepaper, whitepaper2, soyuer2021}, and various mission concepts have  already being discussed \citep[e.g.][]{quest, inpro, whitepaper3, simon}.
The under-exploration of Uranus and Neptune is  unfortunate, as these planets exhibit highly multipolar and nonaxisymmetric magnetic fields \citep{con_ura, con_nep, holme}, are compositionally more diverse than the gas giants \citep{helled, nettel}, and have a significant contrast in their thermal flux. 
More specifically,  Neptune's energy balance (i.e. the ratio of emitted thermal energy to absorbed solar energy) is 1.5--1.75 times that of gas giants, whereas Uranus is almost in equilibrium with the solar flux \citep{pearl1,pearl2}. 

The last two decades have seen dramatic improvements in modelling dynamos of solar system giants \citep{stanley2004,  stanley2006, soderlund2013, jones, wicht}, but our understanding of planetary magnetic fields is still very limited. 
One of the challenges in modelling planetary dynamos is linked to the fact that the magnetic fields are measured external to the planets, whereas the field itself is generated deeper inside the planet.
This inverse source problem makes it difficult to find a unique solution to the dynamo region that generates most of the observed external magnetic fields.
This is especially true for the peculiar magnetic fields of ice giants \citep{krista2020}, due to the relative lack of data for modelling the magnetic fields \citep{holme}, the compositions \citep{helled, nettel}, and the heat transfer mechanisms inside the planets \citep{podolak, vazan}.
Understanding the observed surface magnetic field of Uranus and Neptune does not only involve modelling the dynamo generation region, but also the shallow region quasi-dynamo action, which couples the zonal winds to the background magnetic field.
Investigations of this coupling have been conducted for the gas giants \citep{cao2017, galanti2020}, where it was found that poloidal magnetic field perturbations that are spatially correlated with the zonal flows were possible (with the strength of 0.01 -- 1 per cent of the background field), and that the zonal wind-magnetic field interaction in the semiconducting region of gas giants plays a key role in zonal flow decay.  

We estimate the strength of the shallow layer coupling of the magnetic field to the zonal winds by using the electrical conductivity prescription for H$_2$--He--H$_2$O mixtures provided in \citet{soyuer2020}.
Inspired by \citet{cao2017}, we look for poloidal field perturbations to the background magnetic field induced by this interaction, via a simplified $\alpha$--$\omega$ mechanism.
By compiling sets of zonal wind decay profiles obeying Ohmic dissipation constraints, we estimate the magnitude of poloidal field perturbations for various Uranus models in the literature.
Naturally, our simple estimate is not expected to capture the complete physical picture.
Nevertheless, the goal of this short paper is to draw attention to this mechanism, and show that a mission to Uranus can significantly improve  our understanding of its dynamics, magnetic field and composition.

\section{Uranus Models}
\subsection{Interior structure models}
\begin{figure*}
    \centering
    \includegraphics[width = \textwidth]{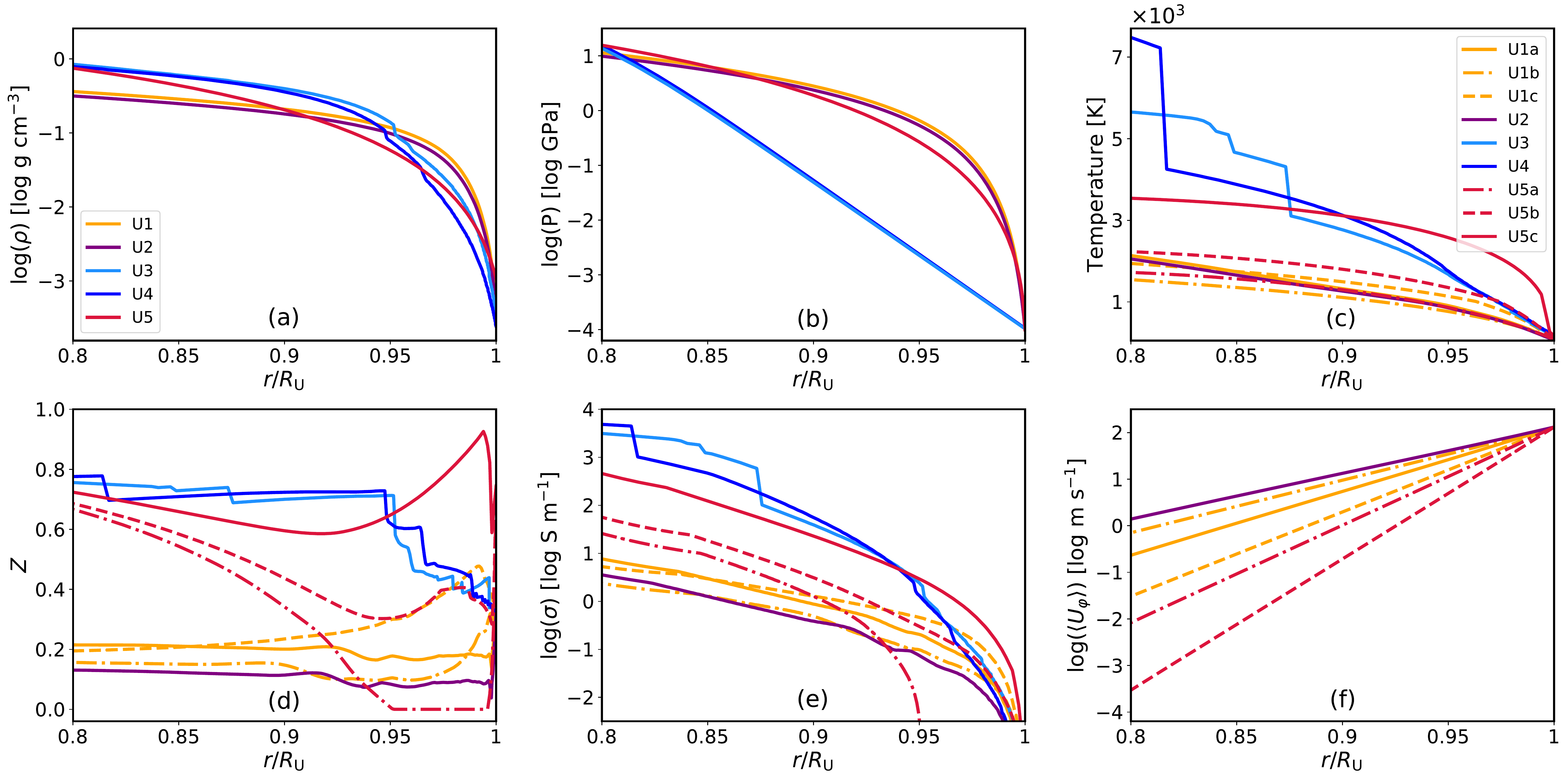}
    \caption{(a): Radial density profiles of various published Uranus structure models  \citep{helled, nettel, podolak, vazan}. (b): Radial pressure profiles of the same models. (c): Various radial temperature profiles corresponding to the interior structure profiles by color. Different linestyles represent alternative temperature profiles for the same colored density/pressure model. 
    (d): Inferred radial metallicity profiles assuming an ideally mixed H$_2$--He--H$_2$O mixture and the aforementioned EOS's.
    (e): Radial electrical conductivity profiles calculated using the interior structure profiles, using the prescription in \citet{soyuer2020} for ionically conducting H$_2$--He--H$_2$O mixtures. (f): The highest allowed rms zonal flow speeds for each interior structure model, evaluated in the aforementioned parameter range. In hot models U3, U4 and U5c, winds need to decay to $\sim$0.1\% of their surface values to admit solutions. Thus, we exclude them from our perturbation analysis and omit them in the  last panel.}
    \label{fig:rm}
\end{figure*}

We consider five different Uranus density/pressure profiles with nine associated temperature profiles taken from \citet{helled, nettel, podolak, vazan} summarized in Table \ref{tab:u_tab}, and explained in detail in \citet{soyuer2020}. 
These models cover a wide range of thermodynamic parameters and also employ different heat transfer mechanisms (convection, double diffusive convection, conduction) and were constructed with various methods and constraints.  Namely; U1a is a 3-layer adiabatic profile by \citet{nettel}, U1b and U1c are the modified double diffusive temperature profiles to U1a by \citet{podolak}. U2 is again a 3-layer adiabatic profile by \citet{nettel}, with a modified rotation period obtained by \citet{helled_shape}. U3 and U4 are non-adiabatic models by \citet{vazan}, evolved using various primordial composition distributions and initial energy budgets, to fit present day Uranus models. U5a, U5b and U5c are empirical density profiles by \citet{helled} with various glued temperature profiles by \citet{podolak}, described in Table \ref{tab:u_tab}.
The top panels in Figure \ref{fig:rm} show the density, pressure, and temperature profiles of the models for the outer 20\% of the planet.
\begin{table*}
	\centering
	\caption{List of the interior structure models of Uranus used in this study. Model numbers correspond to various density/pressure profiles. Letters following the model number (e.g. U1a, U1b, U1c) represent  different temperature profiles for the same density/pressure profile. The original names of the structure models in their respective papers are shown on the right column.}
	\begin{tabular}{l | c c c c c} 
		\hline
		 & Density/Pressure Profile &  Temperature Profile & Rotation Period &  Convective Layers & Original Name  \\
		\hline
		U1a & \citet{nettel} & \citet{nettel} & 17.24h & 1 & U1\\
		U1b & \citet{nettel} & \citet{podolak} & 17.24h &1 & U1 Cold Model\\
		U1c & \citet{nettel} & \citet{podolak} & 17.24h &10$^6$ & U1 Hot Model\\
		U2  & \citet{nettel} & \citet{nettel} & 16.58h &1 &  U2\\
		U3  & \citet{vazan} & \citet{vazan} & 17.24h &-- &  V3\\
		U4  & \citet{vazan} & \citet{vazan} & 17.24h &-- &  V4\\
		U5a & \citet{helled} & \citet{podolak} & 17.24h &1 &  PolyU Cold Model\\
		U5b & \citet{helled} & \citet{podolak} & 17.24h &10$^6$ &  PolyU Hot Model\\
		U5c & \citet{helled} & \citet{podolak} & 17.24h &10$^7$ &  --\\
		\hline
	\end{tabular}
	\label{tab:u_tab}
\end{table*}

For the purpose of this work we assume  the region we are interested in Uranus is composed  of a  H$_2$--He--H$_2$O mixture given by
\begin{equation}
\frac{1}{\rho} = \frac{X}{\rho_\textrm{\tiny H$_2$}} + \frac{Y}{\rho_\textrm{\tiny He}} + \frac{Z}{\rho_\textrm{\tiny H$_2$O}},
\end{equation}
with a protosolar ratio of $X/Y = 0.745/0.255$.
For hydrogen and helium we adopt the equation of state (EOS) developed by \citet{cms} and for water that of \citet{shah}.
The different EOSs are combined using the above isothermal-isobaric ideal volume law, which is a good approximation under the range of conditions explored in the current work for hydrogen--helium \citep{cms}, for water--hydrogen \citep{francois_diss} and for ternary mixtures \citep{francois_HHeZ}. Figure \ref{fig:rm} (d) shows the inferred metallicities from this prescription.

\subsection{Electrical conductivities}

Estimating the ionic conductivity of any H$_2$--He--[liquid ice] mixture is difficult, let alone generalizing the calculation to various mixture parameters.
Arguably, the most notable measurement of electrical conductivity of a mixture resembling the interiors of Uranus and Neptune has been carried out by \citet{nellis_ice}.
In \citet{nellis_ice}, the electrical conductivity of a "synthetic Uranus" mixture composed of water, ammonia and isopropanol (C$_3$H$_8$O) has been measured up to 0.75 Mbar (corresponding to 5000 K in their model).
The experiment shows that the electrical conductivity of this mixture is similar to that of pure water, measured by \citet{hamann, nellis82} for roughly the same regime.
Lately, laser-driven shock-compression experiments have verified the superionic conduction of water ice \citep{millot1}, and ammonia \citep{ammonia} under planetary conditions.

In \citet{soyuer2020} we have developed a model for estimating the ionic conductivity of H$_2$--He--H$_2$O mixtures under planetary conditions. 
This model makes use of the fluctuation-dissipation theorem, where the autocorrelation of microscopic currents are linked to the electric conductivity of ions in a mixture, taking into account the diffusion, electrical charge and number density of various H and O species.
We apply this  prescription to various H$_2$--He--H$_2$O mixtures that fit to Uranus structure models using the EOSs mentioned above. For a detailed description of the ionic conductivity prescription we refer the reader to \citet{soyuer2020}.

Figure \ref{fig:rm} (e) shows the radial electrical conductivity profiles calculated for the Uranus structure models. 
Due to the considerable difference between  interior structure models there is a significant variation in electrical conductivity values (reaching up to three  orders of magnitude at 0.85$R_\mathrm{\scriptscriptstyle{U}}$). 
Hotter models reach higher conductivity levels mainly due to higher metallicities (i.e. higher water concentration) and due to the increase in the fraction of ionized H and O species with temperature.

\subsection{Zonal flow decay}
Uranus exhibits fast zonal winds on the surface with speeds reaching up to roughly 200 m s$^{-1}$ with respect to its rotation. 
The surface wind profile has a retrograde motion around the equator and prograde motion at higher latitudes, fit by
\begin{equation}
v_{\varphi,\mathrm{\scriptscriptstyle{U}}} (r = R_\mathrm{\scriptscriptstyle{U}}, \,\theta) =  170 \times \left(0.6\sin\theta + \sin3\theta\right) \,\,\mathrm{m s^{-1}},
\label{eq:w_ura}
\end{equation}
where $\theta$ is the co-latitude \citep{hammel}.

It is commonly assumed that the surface winds continue downward into the planet, along cylinders parallel to the planetary rotation axis.
There is evidence from gravity data that winds are expected to attenuate with depth in Uranus and Neptune \citep{kaspi2013}. 
Indeed, deep-seated strong winds would violate energy and entropy constraints throughout the planet's interiors due to excess Ohmic dissipation caused by their interaction with the background magnetic field \citep{soyuer2020}. 
Relatively shallow winds also seem to be the case in the gas giants, where estimates based on Juno and Cassini gravity data \citep{kaspi2018,kaspi2020, galanti2020} have similar implications to those based on Ohmic dissipation constraints \citep{liu2008, wicht}. 

We define a general behaviour for an azimuthally symmetric zonal wind profile $U_\varphi (r,\theta, n_i)$ with a radial decay profile $\mathcal{B}(r, n_i)$,  where $n_i$ are any  parameters describing the decay:
\begin{equation}
U_\varphi (r, \theta, n_i) = v_\varphi(\mathcal{A}(r,\theta)) \times \mathcal{B}(r, n_i).
\end{equation}
Here, $\mathcal{A}(r,\theta)$  describes the relationship of any point $(r,\theta)$ inside the planet to the surface wind profile $v_\varphi(r = R_\mathrm{\scriptscriptstyle{U}}, \theta)$.
In the case of winds retaining their surface profile along  lines parallel to the rotation axis, this would be $\mathcal{A}(r,\theta) = \arcsin\left(r \sin\theta / R_\mathrm{\scriptscriptstyle{U}}\right)$.
For zonal wind decay inside the planet, we adopt a simple exponential decay. More sophisticated models for the wind decay such as that in \citet{galanti2020} have been also considered, and have yielded very similar results to those of the exponential decay.
Thus, here we adopt an exponential decay for  simplicity with an e-folding depth $H$:
\begin{equation}
\mathcal{B}(r, H) = \exp\left((r/R_\mathrm{\scriptscriptstyle{U}} -  1)/H \right).
\end{equation}
\subsection{Ohmic dissipation}
The induced electrical currents due to the interaction of the zonal flows with the magnetic field produce Ohmic dissipation given by
\begin{equation}
    P = \int \frac{\bm{j}^2}{\sigma}dV,
    \label{eq:ohm}
\end{equation}
where $\bm{j}$ is the current density given by Ohm's law: $\bm{j} = \sigma(\bm{E} + \bm{U} \times \bm{B})$. 
The prescription for calculating  the current densities are omitted for brevity, but are provided explicitly in \citet{liu_phd, soyuer2020}. 
The current densities can be approximated as
\begin{subequations}
\begin{align}
&j_\theta \approx  \frac{\sigma(r)}{r}\left(\partial_ \theta \int_r^R (\bm{U}_\varphi \times \bm{B}^0_P)_r dr' + r 
(\bm{U}_\varphi \times \bm{B}^0_P)_\theta\right),
\label{eq:jthe}\\
&j_\varphi \approx \frac{\sigma(r)}{r}\partial_\varphi \int_r^R (\bm{U}_\varphi \times \bm{B}^0_P)_r dr',
\label{eq:jphi}
\end{align}
\end{subequations}
where the radial currents $j_r$ are suppressed due to the rapidly varying radial electrical conductivity.

Using equation (\ref{eq:ohm}), we compile sets of zonal wind decay profiles \{$U_\varphi$ | $P(U_\varphi) < P_{\rm lim}$\} for the various Uranus models, where we explore the parameter space  $H \ge 0.01$
with the requirement that the  Ohmic dissipation $P$ remains smaller than an Ohmic dissipation limit $P_{\rm lim}$.
The total Ohmic dissipation can be constrained by the energy or the entropy budget throughout the planetary interior \citep{hewitt, wicht}.
We choose the latter as our limit for two reasons; (i) it does not require the adiabatic cooling to cancel out dissipative heating at each radius, and (ii) it provides a looser constraint compared to the former, allowing us to probe a larger range of flow decay profiles.
We consider the Ohmic dissipation generated above $r_{\scriptscriptstyle } = 0.8 R_{\rm \scriptscriptstyle U}$ and  use the entropy limit
\begin{equation}
    P(U_\varphi)\Big|_{0.8 R_{\rm \scriptscriptstyle U}}^{R_{\rm \scriptscriptstyle U}} \overset{!}{\le} P_{\rm lim} = \left(\frac{T(r = 0.8 R_{\rm \scriptscriptstyle U})}{T_0} - 1\right) L,
\end{equation}
with $L$ being Uranus' surface luminosity and $T_0$ is the temperature at the boundary of the convective envelope, which we set as the 1 bar temperature,  given by the  structure models. Since the Ohmic dissipation per volume associated with the zonal flows is proportional to $p \propto j^2/\sigma \propto U^2_\varphi \sigma$, the contribution to the total built-up Ohmic dissipation starts to diminish after some depth. This occurs due to saturating electrical conductivity and fast zonal flow decay. Therefore, considering the Ohmic dissipation generated above $r = 0.8 R_{\rm \scriptscriptstyle U}$ is an adequate approximation. It also has the added benefit of avoiding the uncertainties in Ohmic dissipation generation arising in the dynamo generation region, and allowing us to probe a greater range of decay profiles due to a less stringent constraint.

It is convenient to describe the zonal wind strength as a function of depth via the rms zonal wind velocity, given by:
\begin{equation}
\langle U_\varphi \rangle (r)  = \left(\frac{1}{2} \int_0^\pi  U_\varphi(r,\theta)^2 \sin{\theta} d\theta\right)^{1/2}.
\label{eq:rms}
\end{equation}
The last panel  in Figure \ref{fig:rm} shows the rms zonal flows that have the highest allowed speeds at 0.8$R_{\rm \scriptscriptstyle U}$ for  Uranus structure models, satisfying the entropy limit.
It is clear from the figure that colder models permit zonal flows that can reach up to 1 m s$^{-1}$ at $r = 0.80 R_{\rm \scriptscriptstyle U}$. For hot models U3, U4 and U5c the zonal winds need to decay to $\sim$0.1\% of their surface values to admit solutions in the Ohmic dissipation framework. 
Note that the plotted zonal wind velocities are the fastest speeds allowed in the Ohmic dissipation framework. Namely, existence of such wind decay profiles would imply that the entropy at $r > 0.8R_{\rm \scriptscriptstyle U}$ is solely generated by the induced currents due to zonal winds.

\begin{figure}
    \centering
    \includegraphics[width = \linewidth]{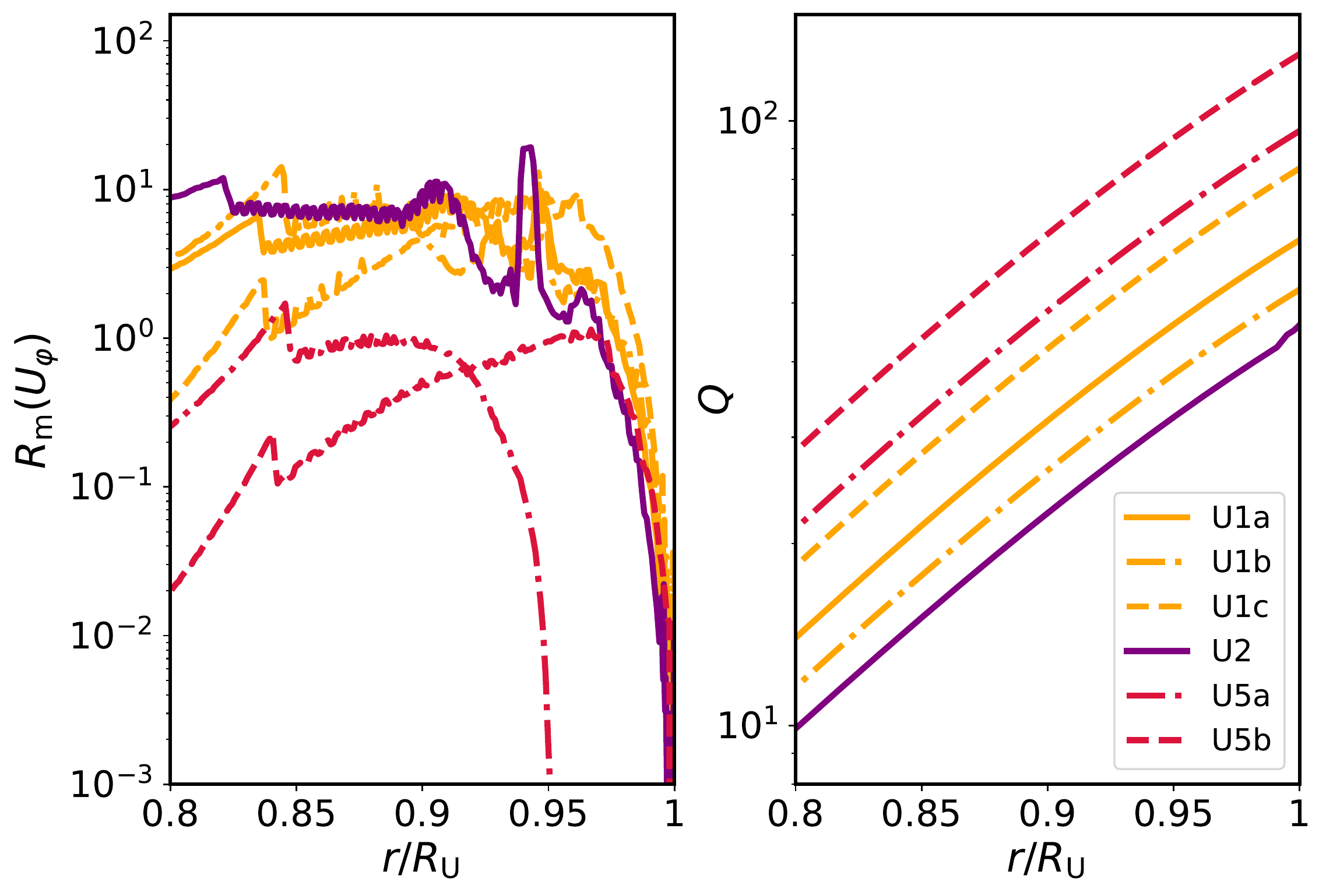} 
    \caption{{Left panel: The magnetic Reynolds number $R_{\rm m}(U_\varphi)$ associated with the extreme zonal flows shown in Figure \ref{fig:rm}.
    Right panel: Ratio $Q$ of toroidal field to poloidal field generation given in equation (\ref{eq:ratio}) for wind decay  corresponding to those presented in Figure \ref{fig:rm}. Although the poloidal field generation is up to two orders of magnitude smaller than the toroidal field generation, it could still have a significant contribution to the total detectable poloidal perturbation $\bm{B}^{\prime}_P$. This is because  the toroidal field $\bm{B}^{\omega}_T$ will cause a poloidal perturbation with a strength of order $B^{\omega}_T R_{\rm m}(\alpha)$, where $R_{\rm m}(\alpha)$ is between $10^{-3}$ and $10^{-2}$ at $r = 0.8 R_{\rm \scriptscriptstyle U}$ for these models.}}
    \label{fig:torpol}
\end{figure}
The inferred magnetic Reynolds number for these zonal flows are shown in the left panel of Figure \ref{fig:torpol}, defined as
\begin{equation}
    R_{\rm m}(U_\varphi) =  \mu_0 \sigma \langle U_\varphi \rangle H_\sigma,
    \label{eq:rm1}
\end{equation}
with $\sigma$ being the electrical conductivity and $H_\sigma = \sigma /\partial_r \sigma$ is  the associated scale height. $R_{\rm m}(U_\varphi)$ essentially describes the zonal wind--background magnetic field coupling in this region.  

\section{Magnetic Field Perturbations}
Planetary magnetic fields $\bm{B}$ are generally decomposed into their poloidal $\bm{B}_P$ and toroidal $\bm{B}_T$ components \citep{backus}, such that
\begin{equation}
    \bm{B} = \bm{B}_P + \bm{B}_T,\, {\rm with} \,\,\,\bm{r} \cdot (\nabla \times\bm{B}_P) = 0 \,\,\,\, {\rm and} \,\,\,\, \bm{r} \cdot \bm{B}_T = 0.
\end{equation}
Uranus' magnetic field is modelled so that the  toroidal component vanishes $\bm{B}_T =0$ external to the planet due to the absence of currents.
At the surface $(r = R_{\rm \scriptscriptstyle U}, \theta, \phi)$, we separate the magnetic field into two parts $\bm{B}_{\rm{surf.}} = \bm{B}^{0}_{P} + \bm{B}^{\prime}_P$, 
where we define $\bm{B}^{0}_{P}$ as the background poloidal field \citep{holme} and $\bm{B}^{\prime}_P$ as the poloidal perturbation to $\bm{B}^{0}_{P}$. We take the surface field as the background field $\bm{B}^{0}_{P} \sim \bm{B}_{\rm{surf.}}$ and extrapolate it downwards. This approximation will not affect the final estimate for the perturbation, since $\bm{B}^{\prime}_P \ll \bm{B}^{0}_{P}$.

We model the generation of the perturbation $\bm{B}^{\prime}_P$ following in the footsteps of \citet{cao2017} with an extra term due to the nonaxisymmetric magnetic field of Uranus:
\begin{enumerate}
\item With increasing depth, zonal winds couple to the background poloidal field via the $\omega$--effect, which induces a field $\bm{B}^{\omega} = \bm{B}^{\omega}_T + \bm{B}^{\omega}_P$. 
\item The poloidal part $\bm{B}^{\omega}_P$ is the first contribution to the detectable field perturbation at the surface $\bm{B}^{\prime}_P$. 
\item The toroidal part $\bm{B}^{\omega}_T$ is then expected to diffuse downwards due to the rapidly increasing electrical conductivity. The $\alpha$--effect acts on the wind induced toroidal field $\bm{B}^{\omega}_T$ and generates a poloidal field $\bm{B}^{\alpha}_P$, the second contribution to $\bm{B}^{\prime}_P$.
\end{enumerate}
Hence, the total poloidal perturbation spatially correlated with the zonal flows and detectable at the surface is given by 
\begin{equation}
\bm{B}^{\prime}_P = \bm{B}^{\omega}_P + \bm{B}^{\alpha}_P.
\end{equation}

In order to estimate the ratio of the toroidal to poloidal contribution of the $\omega$--effect--induced fields, we look at the generative  part (subscript g) of the induction equation
\begin{align}
    (\partial_t &\bm{B}^{\omega})_{\rm g} = \nabla \times (\bm{U} \times \bm{B}^0) = \\ \nonumber
    &- \frac{U_\varphi \partial_\varphi B^0_r}{r \sin{\theta}} \hat{\bm{r}} - \frac{U_\varphi \partial_\varphi B^0_\theta}{r \sin{\theta}} \hat{\bm{\theta}} 
    + \frac{\partial_r(rU_\varphi B^0_r) + \partial_\theta(U_\varphi B^0_\theta)}{r} \hat{\bm{\varphi}},
\end{align}
and set the ratio as the azimuthal field generation to the sum of the radial and latitudinal field generation:
\begin{equation}
    Q := \frac{\bm{B}^{\omega}_T}{\bm{B}^{\omega}_P}(r) \sim \frac{\langle (\partial_t \bm{B}^{\omega})_{{\rm g}} \cdot \hat{\bm{\varphi}} \rangle}{\langle (\partial_t \bm{B}^{\omega})_{{\rm g}} \cdot \hat{\bm{r}}\rangle+ \langle(\partial_t \bm{B}^{\omega})_{{\rm g}}\cdot\hat{\bm{\theta}} \rangle},
    \label{eq:ratio}
\end{equation}
where $\langle \cdot \rangle$ here denotes spherically averaged rms values, i.e. over $\theta$ and $\varphi$ defined analogously to equation (\ref{eq:rms}):
\begin{equation}
    \langle f(\bm{r}) \rangle = \langle f \rangle (r) = \frac{1}{2}\left(\int\limits_0^\pi \int\limits_0^{2\pi} f(\bm{r})^2 d\varphi \sin\theta d\theta\right)^{1/2}.
\end{equation}

Figure \ref{fig:torpol} shows the ratio $Q$  in equation (\ref{eq:ratio}) for the various extreme zonal wind decay profiles given in the bottom center panel of Figure \ref{fig:rm}. 
The field generation in the azimuthal direction is  up to two orders of magnitude larger than in the  poloidal direction. 

An important caveat is that the $\alpha$--effect is not straightforward to model and is the biggest unknown in the prescription. 
The magnetic Reynolds number associated with the $\alpha$--effect is given by
\begin{equation}
    R_{\rm m}(\alpha) = \mu_0 \sigma \alpha H_\sigma,
    \label{eq:rm2}
\end{equation} 
where $\alpha$ is the strength of the effect \citep{cao2017}.
To first order, the strength of the perturbations scale as 
\begin{align}
B^{\omega}_P &=  B^{\omega}  - B^{\omega}_T \sim R_{\rm m}(U_\varphi) B^{0}_{P} - B^{\omega}_T = \frac{R_{\rm m}(U_\varphi) B^{0}_{P}}{1 + Q},  \\[5pt]
B^{\alpha}_P &\sim R_{\rm m}(\alpha) B^{\omega}_T = Q R_{\rm m}(\alpha) B^{\omega}_P,
\end{align}
and  the ratio of the perturbation to the background field is then expressed as
\textbf{\begin{equation}
\frac{B^{\omega}_P + B^{\alpha}_P}{B^0_P} \sim (1 + Q R_{\rm m}(\alpha)) \frac{R_{\rm m}(U_\varphi)}{1 + Q}.
\label{eq:order}
\end{equation}}

Note that, the two magnetic Reynolds numbers are calculated at different depths. $R_{\rm m}(U_\varphi)$ peaks and then saturates (or decreases steadily for slow zonal flows) in the region we have already explored.
The $\alpha$--effect is more pronounced with increasing conductivity, hence $R_{\rm m}(\alpha)$ as well.
In \citet{cao2017}, $\alpha$ is taken to be a scalar functional proportional to the electrical conductivity and having a strength of 0.1 mm s$^{-1}$ at the depth where $\sigma = 10^3$ S m$^{-1}$, corresponding to 10\% of the estimated convective velocity in gas giants at that depth. This is taken to be the base of the semiconducting region in gas giants, i.e., the threshold below which the mean-field induction equation becomes inadequate for describing the behavior of magnetic field generation. 
Using the same $\alpha$--effect strength for our estimations, we get poloidal perturbations that can reach up to
\begin{equation}
\frac{B^{\omega}_P + B^{\alpha}_P}{B^0_P} = \mathcal{O}(0.1)
\end{equation}
for the coldest models (which have flatter zonal wind decay).
Clearly, this estimate cannot capture the complete physical picture. Indeed, as seen in \citet{cao2017} for example, the ratio of the calculated toroidal field perturbation to the background magnetic field is one order of magnitude lower than $R_{\rm m}(U_\varphi)$, when the mean-field equations are solved for an axisymmetric field.
Nevertheless, the detection of zonal wind-induced magnetic field perturbations is possible even for values that are orders of magnitude smaller than this value. Even the   sensitivity of the magnetometer of  \textit{Voyager II}  was of the order of 0.1 nT (roughly a millionth of the surface field strength in ice giants) \citep{holme}.

\section{Discussion and Conclusion}
In this short paper, we have investigated the zonal flow--background magnetic field interaction in Uranus. 
Our findings suggest that:
\begin{enumerate}
    \item Due to their high temperatures in outer regions of hot Uranus models, the  winds need to decay to $\sim$0.1\%  of their surface values in the outer 1\% of Uranus to admit decay solutions. 
    \item Colder models of Uranus do admit flatter decay solutions and  can have wind speeds up to 1 m s$^{-1}$ at $r = 0.80 R_{\rm \scriptscriptstyle U}$ in the extreme. 
    \item Assuming the same $\alpha$--effect strength present at the bottom of the semiconducting region of gas giants, poloidal field perturbations generated by the zonal wind--magnetic field--convection interaction in ice giants can reach up to $\mathcal{O}(0.1)$ of the background field. 
\end{enumerate}
For the last point we repeat the three  important caveats that should be considered. 
First, the scaling given in equation (\ref{eq:order}) is a simplification of the interaction between the zonal wind and magnetic field, and the poloidal perturbation to background field ratio estimated here is roughly an upper limit. 
Second, the uncertainty in the $\alpha$--effect coupling is not trivial. However, via the constraints that will be put on the heat transfer mechanisms after a mission, a better estimate for the $\alpha$--effect can be constructed.
Most importantly, the zonal wind velocities probed here represent the \textit{fastest speeds} allowed in the Ohmic dissipation framework. Such wind decay profiles mean that the entropy at $r > 0.8R_{\rm \scriptscriptstyle U}$ is solely generated by the zonal winds--magnetic field interaction, which is most probably not the case. 

It should be noted that our estimates all use the ionic conductivity estimates from \citet{soyuer2020}, and assume that Uranus is composed of a mixture of H$_2$--He--H$_2$O in the region of interest.
Clearly, the composition of Uranus is more complex and includes other constituents.
However, volatiles like water, ammonia and also similar mixtures are expected to become electrically conducting \citep{nellis82, nellis_ice, millot1, ammonia}, and to have similar conductivities with our estimates. Nevertheless, electrical conductivities of volatile mixtures under planetary conditions requires further research.

Given that region where the perturbations are generated is shallow, high order harmonics might be pronounced at the planetary surface. 
However, in order to get a more complete understanding of  the zonal wind--magnetic field interaction in general, the mean-field induction equations for the toroidal and poloidal field perturbations in the rapidly increasing conductivity region should be modelled and numerically solved.
The nonaxisymmetry of the surface magnetic fields, the uncertainty of convective velocities, and the possible existence of an inner/outer envelope boundary within ice giants makes this topic challenging and future measurements as well as theoretical calculations are required. 
It is clear that there are still many key open questions regarding the nature of Uranus (and Neptune). 
We suggest that a space mission to  Uranus would significantly improve our understanding of its internal structure and composition, the origin of its dynamo, its dynamics and their interplay, and will allow us to test and further develop the ideas presented in this work. 
\section*{Acknowledgements}
We thank the anonymous referee for valuable comments and suggestions.
We thank Hao Cao, David Stevenson, Fran\c{c}ois Soubiran, Yohai Kaspi and Eli Galanti for their feedback on the manuscript. We acknowledge support from SNSF grant \texttt{\detokenize{200020_188460}} and the National Centre for Competence in Research ‘PlanetS’ supported by SNSF. 
\section*{Data Availability}
Models in this work are those used in \citet{soyuer2020} and will be shared on request to the corresponding author(s) with permission.


\bibliographystyle{mnras}
\bibliography{example} 



\bsp	
\label{lastpage}
\end{document}